\documentclass[conference]{IEEEtran}
\usepackage{cite}
\usepackage{amsmath,amssymb,amsfonts}
\usepackage{algorithmic}
\usepackage{graphicx}
\usepackage{textcomp}
\usepackage{xcolor}
\usepackage{comment}
\usepackage{url}
\usepackage{enumerate}
\usepackage[caption=false,font=footnotesize]{subfig}
\usepackage{multirow}
\def\BibTeX{{\rm B\kern-.05em{\sc i\kern-.025em b}\kern-.08em
    T\kern-.1667em\lower.7ex\hbox{E}\kern-.125emX}}

\makeatletter
\def\figcaption{\def\@captype{figure}\caption}
\makeatother

\begin{document}

\title{Follower--Followee Ratio Category and User Vector for Analyzing Following Behavior}

\author{%
\IEEEauthorblockN{Hayato Oshimo\IEEEauthorrefmark{1}, Shiori Hironaka\IEEEauthorrefmark{1}, Mitsuo Yoshida\IEEEauthorrefmark{2}, and Kyoji Umemura\IEEEauthorrefmark{1}}
\IEEEauthorblockA{\IEEEauthorrefmark{1}Department of Computer Science and Engineering\\
Toyohashi University of Technology \\
Aichi, Japan\\
Email: oshimo.hayato.zk@tut.jp, hironaka.shiori.ru@tut.jp, umemura@tut.jp
}
\IEEEauthorblockA{\IEEEauthorrefmark{2}Faculty of Business Sciences\\
University of Tsukuba\\
Tokyo, Japan\\
Email: mitsuo@gssm.otsuka.tsukuba.ac.jp}
}

\maketitle

\begin{abstract}
Analyzing following behavior is important in many applications.
Following behavior may depend on the main intention of the follower.
Users may either follow their friends or they may follow celebrities to know more about them.
It is difficult to estimate users' intention from their following relationships.
In this paper, we propose an approach to analyze following relationships.
First, we investigated the similarity between users.
Similar followers and followees are likely to be friends.
However, when the follower and followee are not similar, it is likely that follower seeks to obtain more information on the followee.
Second, we categorized users by the network structure.
We then proposed analysis of following behavior based on similarity and category of users estimated from tweets and user data. 
We confirmed the feasibility of the proposed method through experiments.
Finally, we examined users in different categories and analyzed their following behavior.
\end{abstract}

\vspace{\baselineskip}
\renewcommand\IEEEkeywordsname{Keywords}
\begin{IEEEkeywords}
    \textit{User analysis, User embeddings, Network science, Twitter, Following behavior}
\end{IEEEkeywords}

\section{Introduction}
Twitter is a social media platform where people post short messages called tweets and communicate with each other.
Twitter users \textit{follow} other users by subscribing to their tweets.
Twitter users can follow without the permission of the targeted user; thus, the following relationship is directed.

Analyzing following behavior is important in many applications, such as friend recommendations~\cite{Barbieri2014} or information diffusion analysis~\cite{Guille2012}.
Users' following behavior depends on their intention.
There are various intentions on the following links~\cite{Takemura2015}.
It is difficult to classify these links because it is hard to collect data indicating the intentions of the links.

We assume that different categories of users have different preferences for whom to follow.
For example, users that are willing to learn more about celebrities follow them.
We classified users by the follower--followee ratio, which is the ratio of the number of followees to the number of followers.
The follower--followee ratio has been used to analyze social media users~\cite{Yan2018}.

We analyzed the preferences of users followers based on the user category and topical similarity.
Topical similarity reflects the similarity between the users' tweets.
The user category was defined using the follower--followee ratio, which reflects the user's characteristics.
First, we confirmed the feasibility of the computed topical similarity.
Then, we confirmed the feasibility of the category using topical similarity.
We found that the following behavior described based on the topical similarity between the follower and followee provided a reasonable explanation for the following relation among users in different categories.
This suggests that both category and topical similarity are useful for analyzing following behavior.

\section{Related Work}
\subsection{User Categories on Social Media}

Java et al.~\cite{Java2007} considered that Twitter users can mainly be categorized as Information Source, Friends, and Information Seeker, based on their link structure.
Yan et al.~\cite{Yan2018} used the follower--followee ratio to determine user characteristics on ResearchGate, a social media platform for scientists and researchers.
ResearchGate users can share their research papers and follow other researchers.
Yan et al. adopted the user categories proposed by Java et al.\ and classified users based on the follower--followee ratio.

Other researchers have performed classification without using link structures, using other methods such as classifying users into five types based on social theory~\cite{Priante2016} and estimating Big Five personalities from user profiles~\cite{Golbeck2011}.
These classifications require training data collected through surveys or crowdsourcing.

We classified Twitter users into four categories based on their follower--followee ratio.
The categories of Information Source and Information Seeker were the same as in previous studies~\cite{Java2007,Yan2018}.
In addition, we divided the Friends category into two groups, according to whether the follower--followee ratio was greater than 1.
We assumed that more general users would have a smaller number of followers than followees and would exhabit different characteristics.

\subsection{Purpose and Intention of Following}

Following behavior depends on the purpose of the following, which relates to edge types.
Barbieri et al.~\cite{Barbieri2014} proposed a user recommendation method based on whether the edge is topical or social.
Komori et al.~\cite{Komori2018} analyzed following relationships by classifying them into virtual and real friendships.
Takemura et al.~\cite{Takemura2015} classified following relationships into eight types, combining three axes: user-orientation, content-orientation, and mutuality.
These researchers collected data for each following relationships by using surveys to build a classification model.
However, it is difficult to collect training data on individual following relationships automatically.
Yamaguchi et al.~\cite{Yamaguchi2015} proposed a method to explain the reason for following through coupled tensor analysis using tagging action (add users to the lists).
However, only a few users use the list feature on Twitter.

We consider that different categories of users tend to have different main purposes for following.
We simply classified user categories by follower--followee ratio and analyzed the following behavior according to user category.

\subsection{Homophily on Social Media}

Homophily is a phenomenon where users tend to be friends with similar people~\cite{McPherson2001}.
Various types of homophily have been observed on the online social graph~\cite{Bisgin2010,Zamal2012,Pan2019}.
In a previous study~\cite{Weng2010}, topical homophily was reported based on the user's topics of interest recognized from tweets using latent dirichlet allocation (LDA) and the following relationship, and the authors concluded that the topics of users with following relationships are similar.
We also focused on the topical homophily of the users' tweet content.

Homophily relates to network structure.
The follower--followee ratio and homophily of various attributes have been investigated~\cite{Choudhury2011}.
Homophily is an important assumption in network-based user attribute estimation.
Hironaka et al.~\cite{Hironaka2021} examined the relationship between the follower--followee ratio and location homophily using home location estimation.
Based on the data of the countries that are the top-10 users of Twitter, they reported that the follower--followee ratio contributes to the estimation performance.
In this study, we examined the relationship between topical homophily and follower--followee ratio.

\section{Data Collection}
First, we randomly extracted users for analysis using Twitter API.
Then, we collected data on their followees and followers.
In addition, we collected their tweets to calculate topical homophily.

We collected English tweets from July 11 to July 17, 2021, using Twitter Streaming API\footnote{\url{https://developer.twitter.com/en/docs/twitter-api/v1/tweets/filter-realtime/api-reference/post-statuses-filter} (viewed 2022-06-10)}.
We randomly selected 50,000 unique users who tweeted at least once in this period.

Next, we collected followees and followers data using API\footnote{\url{https://developer.twitter.com/en/docs/twitter-api/v1/accounts-and-users/follow-search-get-users/api-reference/get-followers-ids} and \url{https://developer.twitter.com/en/docs/twitter-api/v1/accounts-and-users/follow-search-get-users/api-reference/get-friends-ids} (viewed 2022-06-10)}.
We also collected the latest 3200 tweets using API\footnote{\url{https://developer.twitter.com/en/docs/twitter-api/v1/tweets/timelines/api-reference/get-statuses-user_timeline} (viewed 2022-06-10)}.
If a user had posted less than 3200 tweets, we collected as many as possible.
As a result, 48,881 user timelines were collected.

In the analysis, we used the data of 48,829 users, that is, the users whose tweet and follower data that we could successfully collect.
We detected 59,778 following relationships among them.

\section{User Classification and User Vector}
\begin{figure}[tp]
    \centering
    \includegraphics[width=0.8\linewidth]{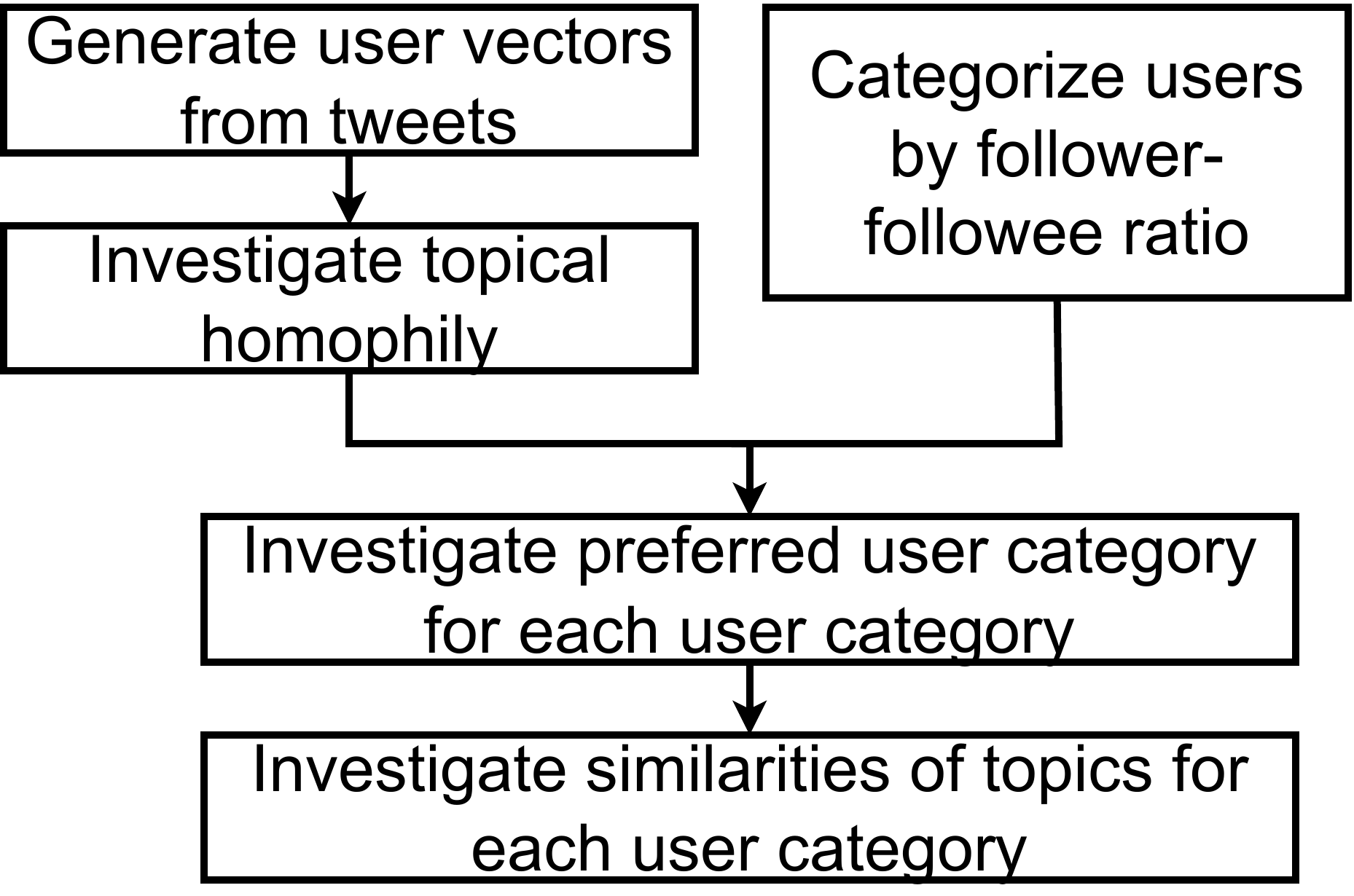}
    \caption{Research workflow}
    \label{fig:workflow}
\end{figure}

In this study, we analyzed users' following behavior based on user category and topical homophily.
We analyzed the following workflow showed in Figure~\ref{fig:workflow}.

First, we explain the follower--followee ratio to classify users and then describe the classification method.
Second, we define the user vector for calculating the topical homophily and then describe the calculation method of topical homophily.

\subsection{Follower--Followee Ratio}

The follower--followee ratio is the ratio of the number of followees $N_{followee}$ to the number of followers $N_{follower}$, as defined in Equation~\eqref{equ:follow_ratio}.
\begin{equation}
    \mbox{follower--followee ratio} = \frac{N_{followee} + 1}{N_{follower} + 1} \label{equ:follow_ratio}
\end{equation}

In Equation~\eqref{equ:follow_ratio}, we add 1 to the denominator to avoid devision by zero and to the numerator to guarantee that the ratio of a user with equal number of followees and followers become 1.

\begin{figure}[t]
    \centering
    \subfloat[High (Information Seeker)\label{fig:high_follow_ratio}]{
        \includegraphics[width=0.45\linewidth]{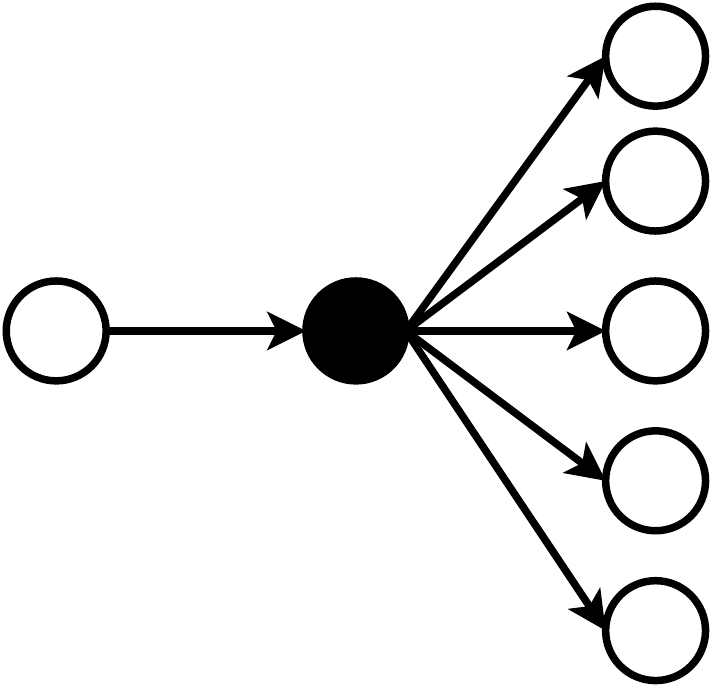}
    }
    \subfloat[Low (Information Source)\label{fig:low_follow_ratio}]{
        \includegraphics[width=0.45\linewidth]{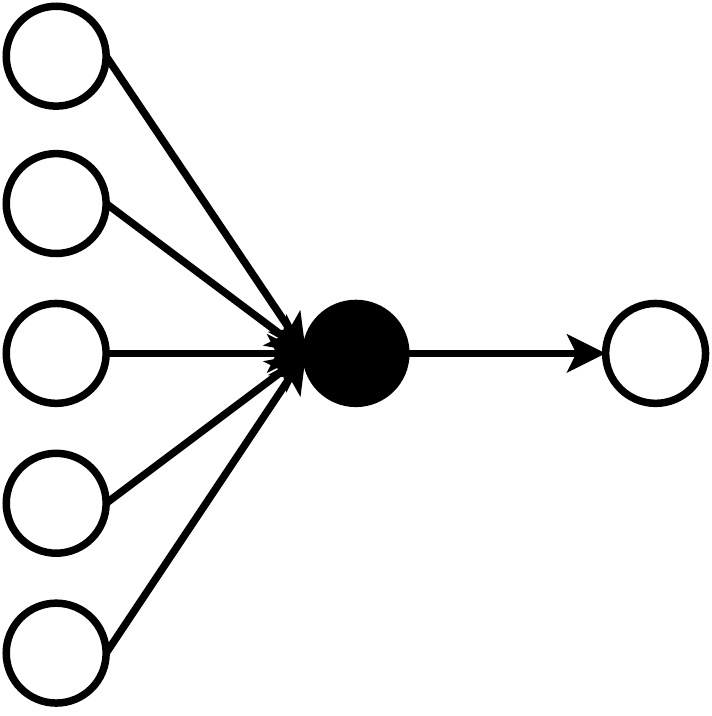}
    }
    \caption{Examples of users with high and low follower--followee ratio.}
\end{figure}

Figure~\ref{fig:high_follow_ratio} and~\ref{fig:low_follow_ratio}, respectively, show examples of users with high and low follower--followee ratios.
Users with a high follower--followee ratio are those whose number of followees $N_{followee}$ is signigicantly outnumbered by the number of their followers $N_{follower}$.
The reverse is the case for users with a low follower--followee ratio.

\subsection{User Classification Using Follower--Followee Ratio}

In this study, we classify users into four categories, A through D, according to the follower--followee ratio.
Category A (Information Seeker) represents users with a follower--followee ratio of 2.0 or higher,
B (Friend) represents users with a ratio between 1.0 and 1.25,
C (Friend Hub) represents users with a ratio between 0.8 and 1.0,
and D (Information Source) represents users with a ratio of 0.5 or lower.
In this study, note that users with a ratio between 1.25 and 2.0 or between 0.5 and 0.8, are not included in any category.
The thresholds were decided based on Hironaka's report~\cite{Hironaka2021icaicta} that the location similarity of users whose follower--followee ratio was 0.5, 1.0, and 2.0 varied.

The number of users in each category is shown in Table~\ref{tab:class_users}.
The average numbers of followees and followers are also shown in the same table.
We obtained these numbers from calculation using the following data.
We used only the following relationships between the users who were included in any categories.
Out of 48,829 users, 44,402 were classified.
Category B contains the most users.
Based on the average number of degrees, the values of Categories B and C vary significantly.
The users in Category C have high numbers of followees and followers.
Hence, we call Category C the Friend Hub.

\begin{table}[t]
    \centering
    \caption{Number of users in each category.}
    \label{tab:class_users}
    \begin{tabular}{|c|c|c|c|c|}\hline
        &A&B&C&D\\\hline
        Number of users&\multicolumn{1}{|r|}{11404}&\multicolumn{1}{|r|}{28616}&\multicolumn{1}{|r|}{671}&\multicolumn{1}{|r|}{3711}\\\hline
        The average number of followees&\multicolumn{1}{|r|}{2.57}&\multicolumn{1}{|r|}{0.79}&\multicolumn{1}{|r|}{7.92}&\multicolumn{1}{|r|}{0.64}\\\hline
        The average number of followers&\multicolumn{1}{|r|}{0.26}&\multicolumn{1}{|r|}{0.74}&\multicolumn{1}{|r|}{8.90}&\multicolumn{1}{|r|}{7.98}\\\hline
    \end{tabular}
\end{table}

\subsection{User Vector}

We construct user vectors to calculate topical homophily between users.
In this paper, topical homophily between users is the similarity of the user's topics of interest recognized from tweets.
A user vector is a document vector constructed by considering a user's tweets as a single document.
Our method is based on Mochihashi's method~\cite{Mochihashi2021} to find researchers from search queries by obtaining a vector representation of papers and researchers.
His method, which is built on Levy's method~\cite{Levy2014}, creates a shifted positive pointwise mutual information matrix (SPPMI) of papers (documents) and performs matrix factorization.
This method is more scalable than neural embedding models like Doc2Vec~\cite{Le2014}.

User vectors are computed as follows:
\begin{enumerate}[1.]
    \item Divide users into sample users and out-of-sample users.
    \item Count word frequencies in tweets.
    \item Select word vocabulary.
    \item Compute user vectors of sample users.
    \item Compute user vectors of out-of-sample users.
\end{enumerate}

Step 1:
To efficiently compute user vectors for a large number of users, we divide users into sample users and out-of-sample users.
We randomly select 10,000 users as sample users from the 48,881 users whose tweets were collected.
The remaining users are out-of-sample users.

Step 2: We count the frequencies of the words in the tweets of each user.
Each tweet text is tokenized using NLTK\footnote{\url{https://www.nltk.org/} (viewed 2022-06-27)}.

Step 3: We select word vocabulary to compute user vectors based on word frequencies.
The word vocabulary is defined as the top 10,000 most-frequent words in all tweets.

Step 4: We construct matrix $M$ from its element $m_{u, w}$ based on the frequencies of the words:
\begin{align}
    \begin{split}
        m_{u, w} &= \max{(\log{\frac{\hat{P}(u, w)}{\hat{P}(u)\hat{P}(w)}} - \log{k}, 0)}\\
        &= \max{(\log{\frac{\hat{P}(w|u)}{\hat{P}(w)}} - \log{k}, 0)}\\
        &= \max{(\log{\frac{\#(u, w) \times D}{\#(u) \times \#(w)}} - \log{k}, 0)}
    \end{split}
\end{align}
where $\#(u, w)$ is the number of occurrences of the word $w$ in the tweet of user $u$,
$\#(u)$ is the total number of words in the vocabulary in user $u$'s tweets,
$\#(w)$ is the number of occurrences of the word $w$ in all the tweets of sample users,
D is the total number of occurrences of the vocabulary words in the all sample users' tweets,
and $k$ is a value corresponding to the number of negative samples on Word2Vec~\cite{Mikolov2013}.
We use $k=1$, which is reported as the best value in the previous study~\cite{Levy2014}.

Next, we compute the user vectors by truncated singular value decomposition (SVD) of the matrix $M$.
We compute matrix $U$ and $W$ using the top $n$ singular values.
The value $n$ should be searched for each task.
In this study, we use $n=200$.
User vectors are computed by the following factorization:
\begin{align}
    \begin{split}
        M \simeq A \Sigma B^T &= UW^T\\
        U = A \sqrt{\Sigma}, \qquad W &= B \sqrt{\Sigma} 
    \end{split}
\end{align}
where $A$ is a $(\text{the number of sample users}) \times n$ unitary matrix,
$B$ is a $(\text{the number of vocabulary words}) \times n$ unitary matrix,
$\Sigma$ is a diagonal matrix of the top $n$ singular values.
We use the rows of matrix $U$ as the user vectors of sample users.

Step 5: We compute user vectors of the out-of-sample users using matrix $W$ computed in Step 4.
First, we generate a vector $\vec{m}_{u*}$ of the SPPMI between out-of-sample user $u*$ and vocabulary words.
This is the same procedure as the calculation of sample users.
\begin{align}
    \begin{split}
        &\vec{m}_{u*} = (\cdots, \max{(\log{\frac{\hat{P}(w|u*)}{\hat{P}(w)}} - \log{k}, 0)},\cdots)\\
        &\hat{P}(w|u*) = \frac{\#(u*, w)}{\#(u*)}, \qquad \hat{P}(w) = \frac{\#(w)}{D}
    \end{split}
\end{align}

Next, we compute user vector $\vec{u}*$ of out-of-sample user $u*$ using $\vec{m}_{u*}$ and $W$.
\begin{align}
    \vec{m}_{u*} &= \vec{u*} W^T\notag\\
    \vec{m}_{u*}^T &= W \vec{u*}^T\notag\\
    W^T \vec{m}_{u*}^T &= W^T W {\vec{u}*}^T\notag\\
    (W^T W)^{-1} W^T \vec{m}_{u*}^T &= {\vec{u}*}^T\notag\\
    \vec{u}*^T &= (W^T W)^{-1} W^T \vec{m}_{u*}^T
\end{align}
Thus, an approximation vector is computed considering that an out-of-sample user was included in the sample users.
It is not necessary to calculate $(W^TW)^{-1}W^T$ repeatedly; therefore the user vector of the out-of-sample user can be calculated efficiently.

Note that, in the computation of the user vector in this study, retweet is considered the same as a normal tweet.
The part of the original tweet of the quote retweet is excluded.
We assume that retweet signaled agreement with the original tweet.
In future studies, the processing of retweets and quote retweets may need to be reconsidered.

\subsection{Definition of Similarity Between Users}

The strength of topical homophily is considered to indicate similarity between users.
To measure it, we use cosine similarity of user vectors as defined by the following equation for user vectors $\vec{a}$ and $\vec{b}$:
\begin{align}
    \cos{(\vec{a}, \vec{b})} = \frac{\vec{a} \cdot \vec{b}}{||\vec{a}|| ||\vec{b}||}
\end{align}
The cosine similarity become one when the vectors are unidirectional and zero when they are orthogonal.
A large cosine similarity indicate high similarity between users.

In this study, we compute user vectors based on users' tweets through matrix factorization.
We confirm whether topical homophily can be observed from the computed user vectors.
If the computed user vectors reflect users' interest, 
we observe high topical similarity between users with following relationships, as in the previous report~\cite{Weng2010}.

Upon comparison, the average similarity between users with following relationship, at 0.4444, was found to be higher than the one between users without (0.0383).
We thus confirmed that topical homophily can be inferred from the computed user vectors.

\section{Analysis}
We analyzed the following preferences according to the four user categories from two perspectives: user category and topical homophily.
First, we compared the number of following edges between user categories.
Then, we investigated the preferred user categories for each user category.
Next, we compared the topical homophily by each edge and investigated whether the users in each category preferred users who posted similar tweets.

Table~\ref{tab:user_pairs} shows the number of following relationships across categories.
The users in Category A (Information Seeker) tended to follow users in Category D (Information Source), indicating that Information Seekers preferred to follow Information Sources.
The users in Category B (Friend), C (Friend Hub), and D (Information Seeker) tend to follow users in the same category, indicating that they preferred users in the same category.

\begin{table}[t]
    \centering
    \caption{Number of following relationships between categories}
    \label{tab:user_pairs}
    \begin{tabular}{|c|c|c|c|c|}\hline
        \multirow{2}{*}{row follows column} & Information & \multirow{2}{*}{Friend} & Friend & Information \\
        & Seeker && Hub & Source \\\hline
        Information& \multicolumn{1}{|r|}{\multirow{2}{*}{799}} & \multicolumn{1}{|r|}{\multirow{2}{*}{2822}} & \multicolumn{1}{|r|}{\multirow{2}{*}{1273}} & \multicolumn{1}{|r|}{\multirow{2}{*}{24457}} \\
        Seeker&&&&\\\hline
        \multirow{2}{*}{Friend}& \multicolumn{1}{|r|}{\multirow{2}{*}{1564}} & \multicolumn{1}{|r|}{\multirow{2}{*}{14338}} & \multicolumn{1}{|r|}{\multirow{2}{*}{3488}} & \multicolumn{1}{|r|}{\multirow{2}{*}{3270}} \\
        &&&&\\\hline
        \multirow{2}{*}{Friend Hub}& \multicolumn{1}{|r|}{\multirow{2}{*}{473}} & \multicolumn{1}{|r|}{\multirow{2}{*}{3299}} & \multicolumn{1}{|r|}{\multirow{2}{*}{1021}} & \multicolumn{1}{|r|}{\multirow{2}{*}{526}} \\
        &&&&\\\hline
        Information& \multicolumn{1}{|r|}{\multirow{2}{*}{199}} & \multicolumn{1}{|r|}{\multirow{2}{*}{692}} & \multicolumn{1}{|r|}{\multirow{2}{*}{196}} & \multicolumn{1}{|r|}{\multirow{2}{*}{1361}} \\
        Source&&&&\\\hline
    \end{tabular}\\
\end{table}

\begin{figure*}[tp]
    \centering
    \subfloat[\centering Information Seeker (A) to\newline Information Seeker (A)\label{fig:hist_A2A}]{
        \includegraphics[width=0.24\linewidth]{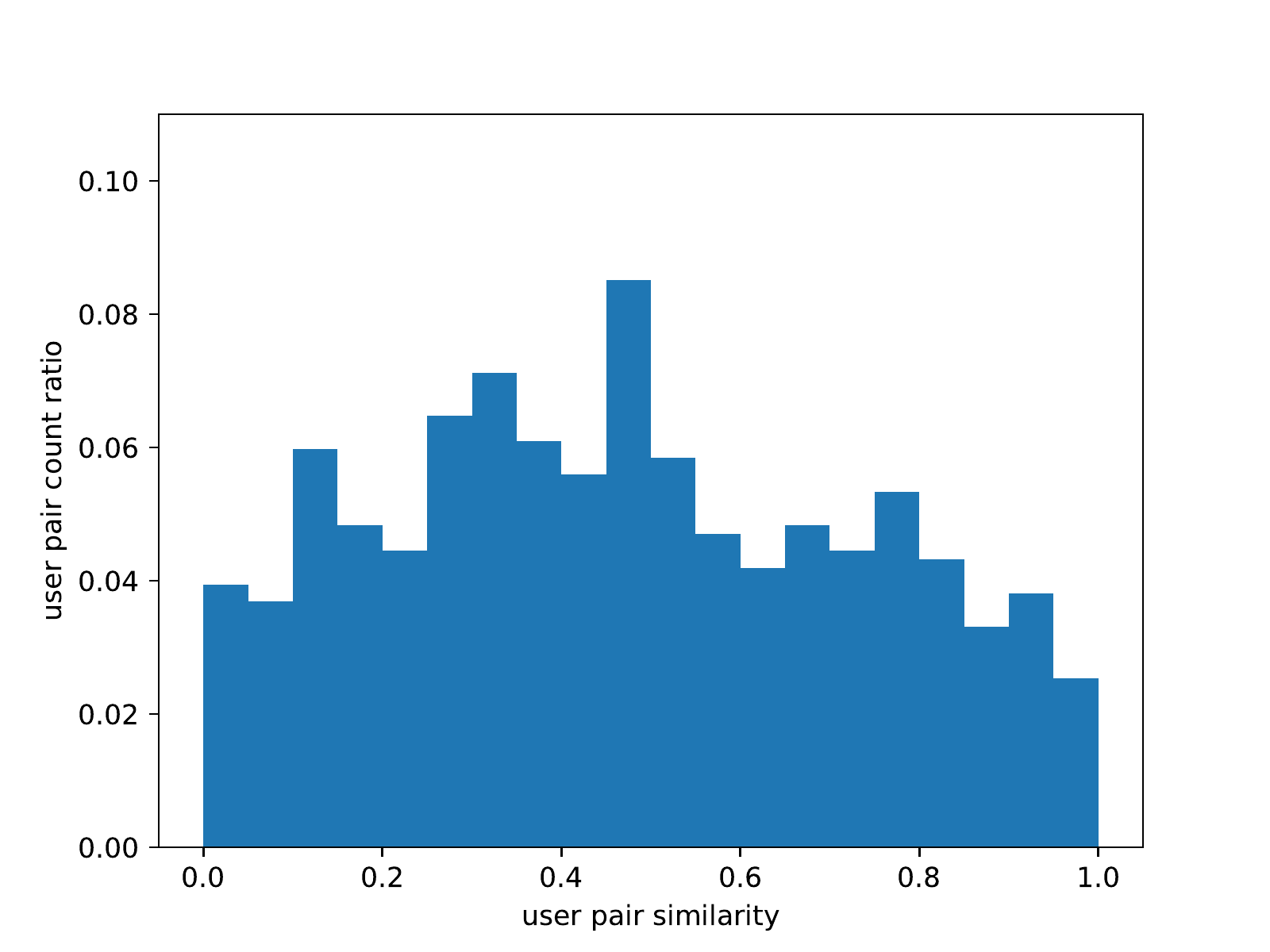}
    }
    \subfloat[\centering Information Seeker (A) to\newline Friend (B)\label{fig:hist_A2B}]{
        \includegraphics[width=0.24\linewidth]{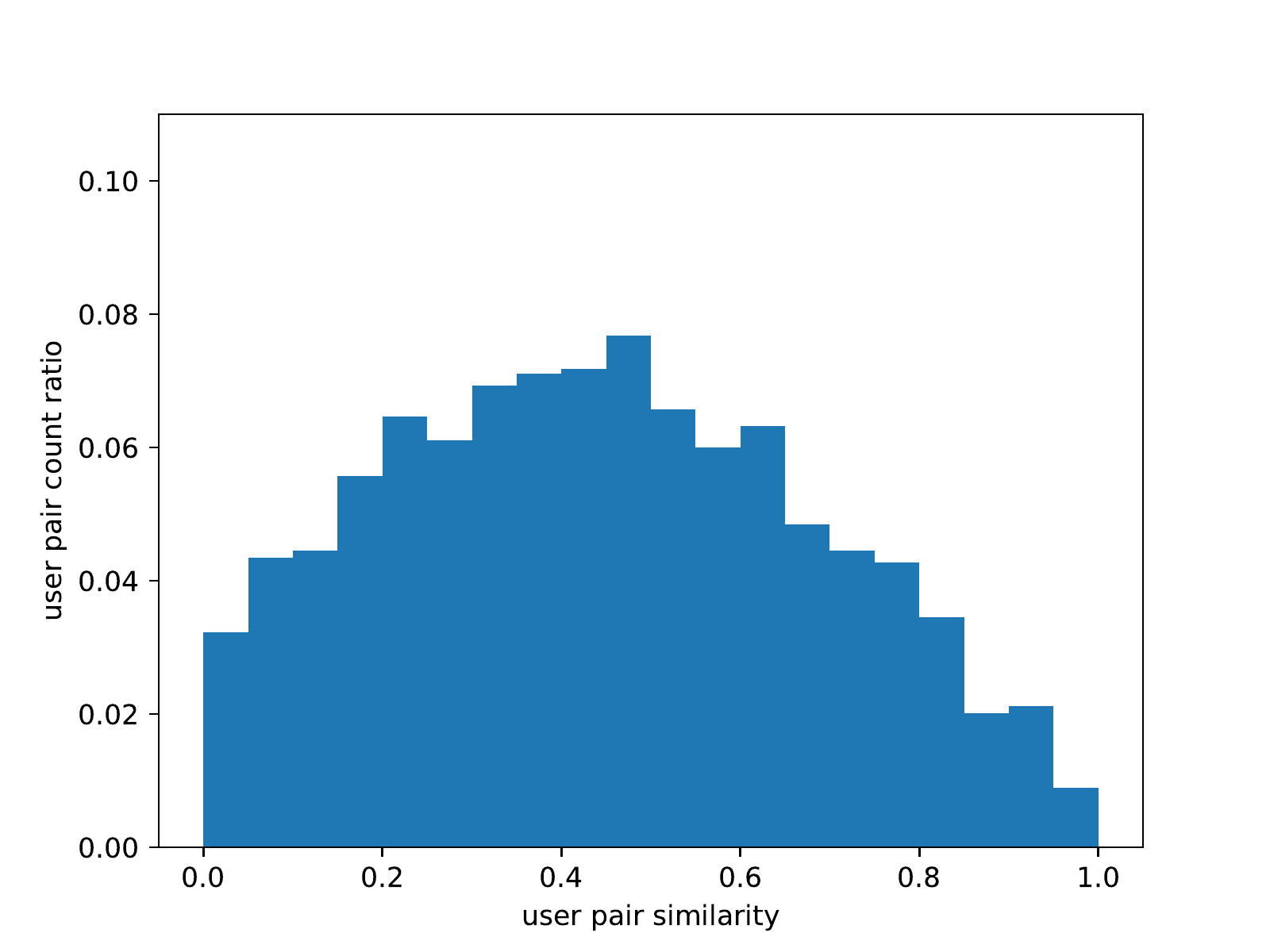}
    }
    \subfloat[\centering Information Seeker (A) to\newline Friend Hub (C)\label{fig:hist_A2C}]{
        \includegraphics[width=0.24\linewidth]{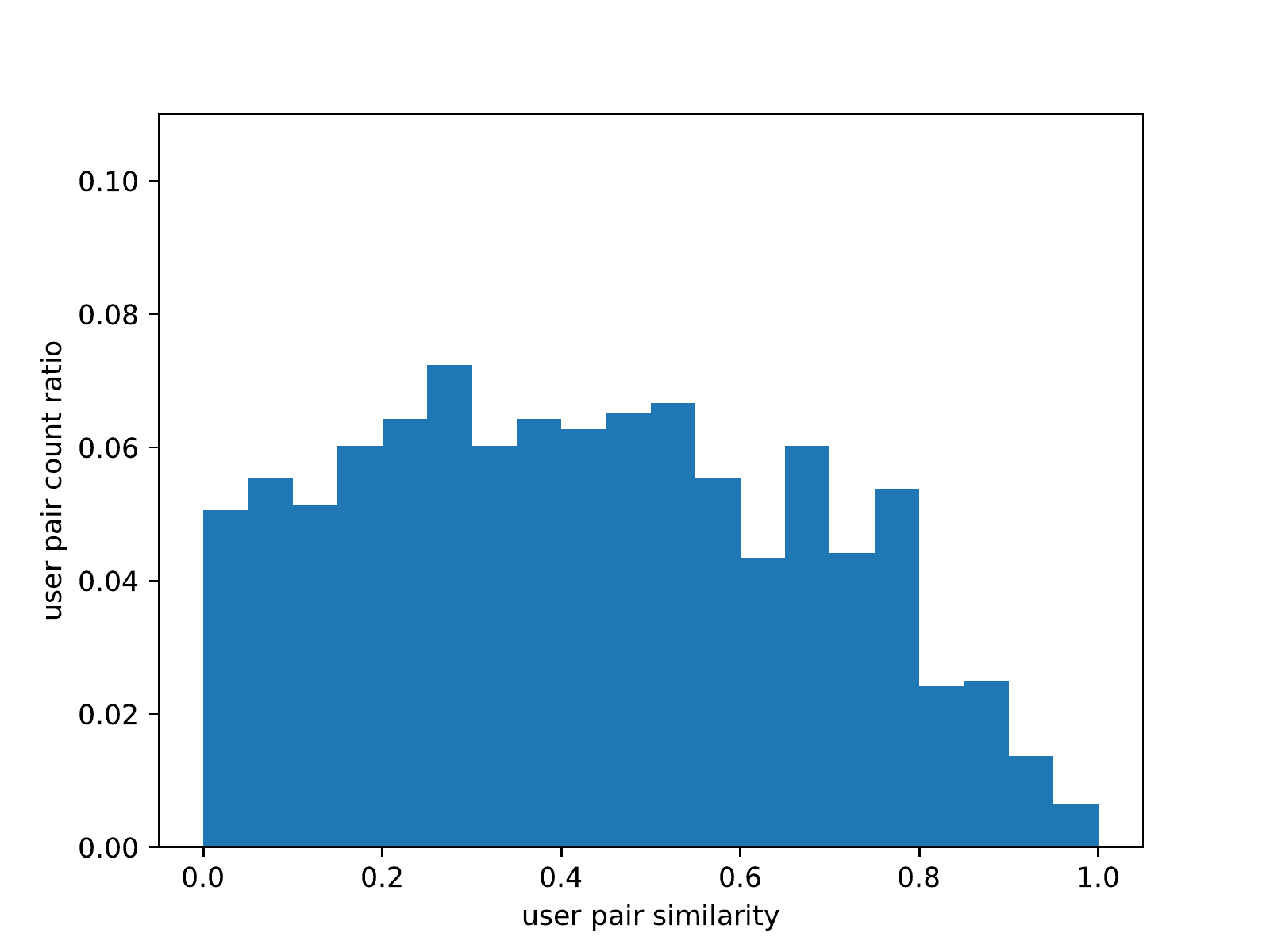}
    }
    \subfloat[\centering Information Seeker (A) to\newline Information Source (D)\label{fig:hist_A2D}]{
        \includegraphics[width=0.24\linewidth]{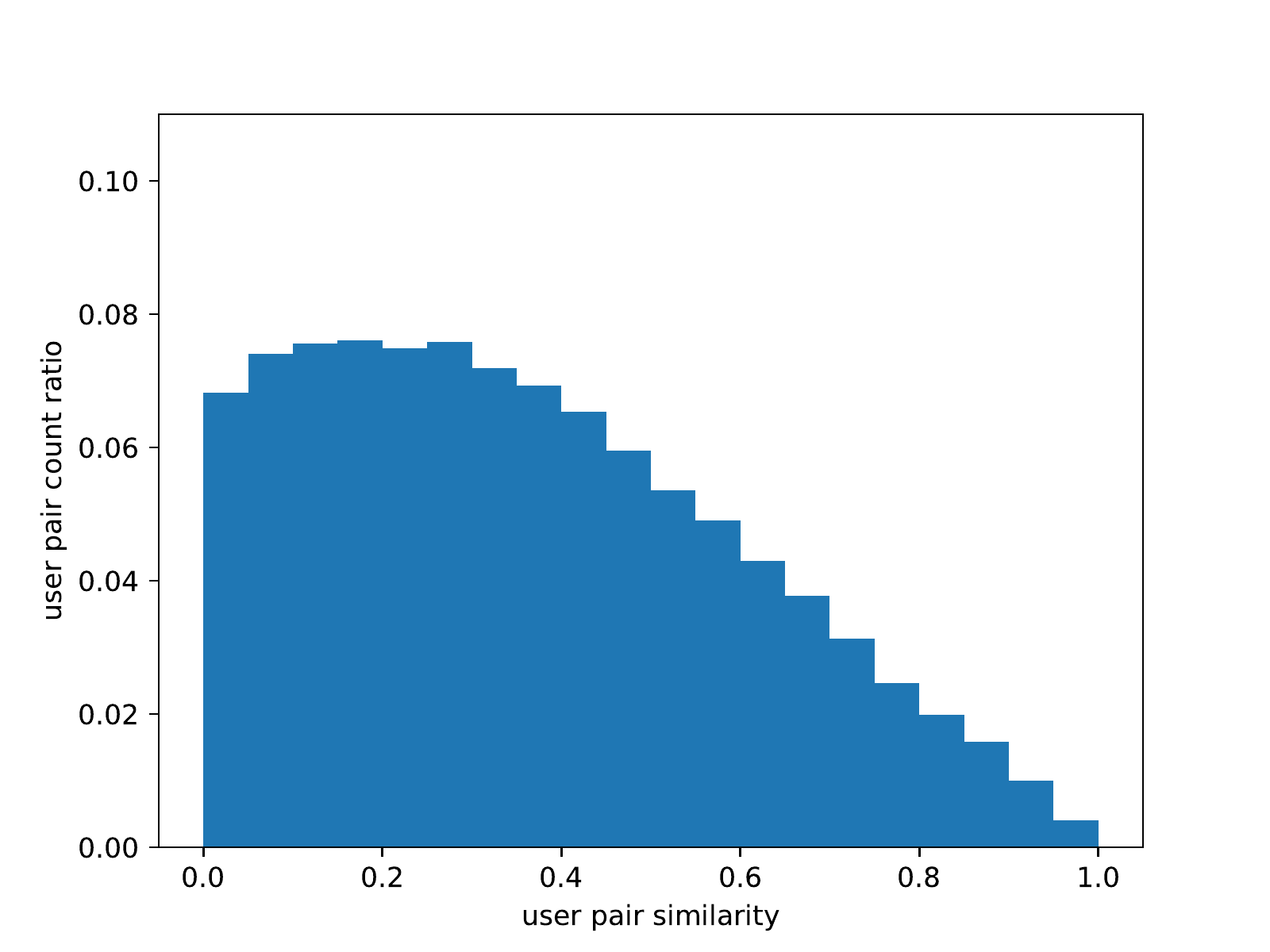}
    }\\
    \subfloat[\centering Friend (B) to\newline Information Seeker (A)\label{fig:hist_B2A}]{
        \includegraphics[width=0.24\linewidth]{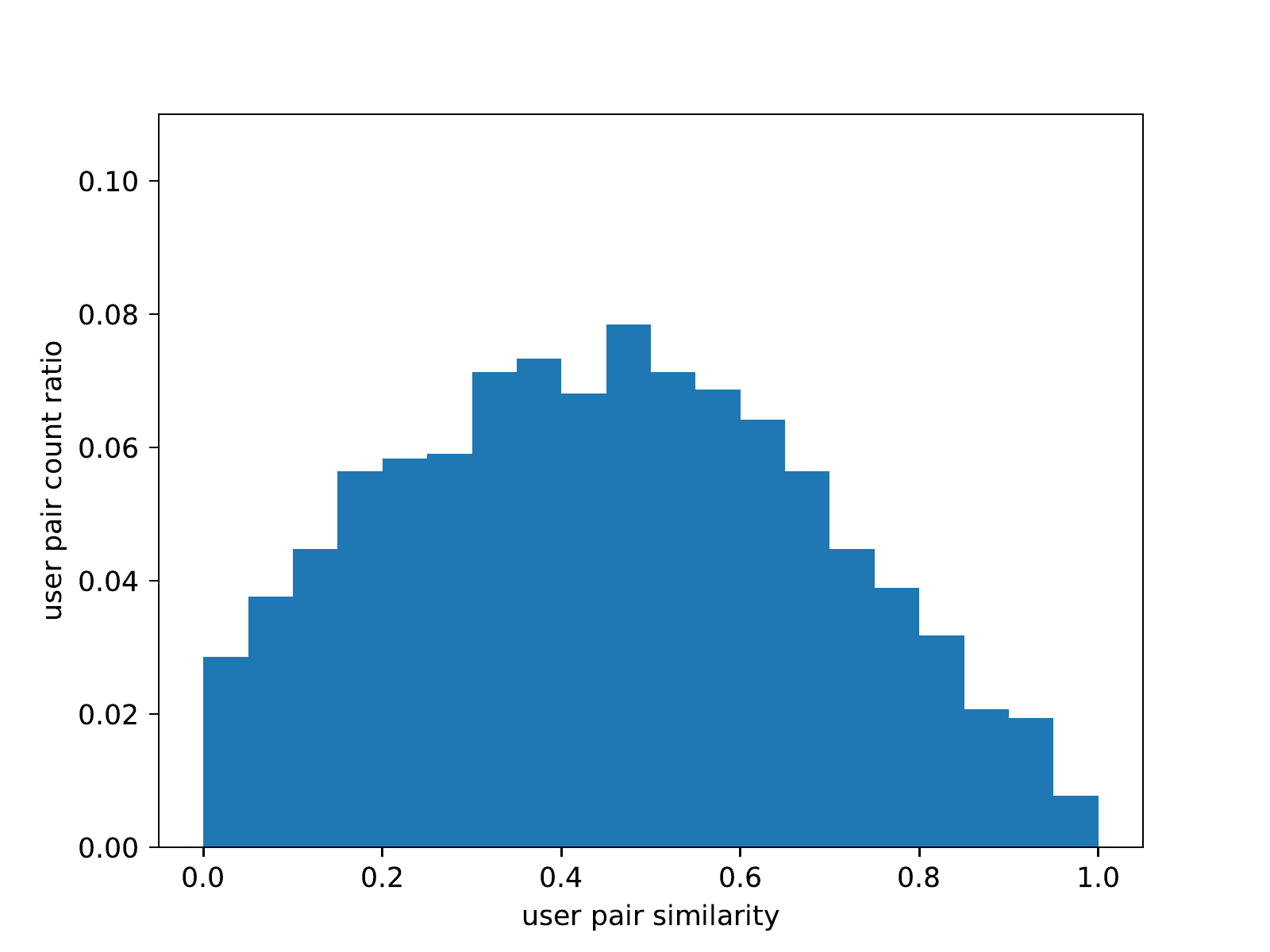}
    }
    \subfloat[\centering Friend (B) to Friend (B)\label{fig:hist_B2B}]{
        \includegraphics[width=0.24\linewidth]{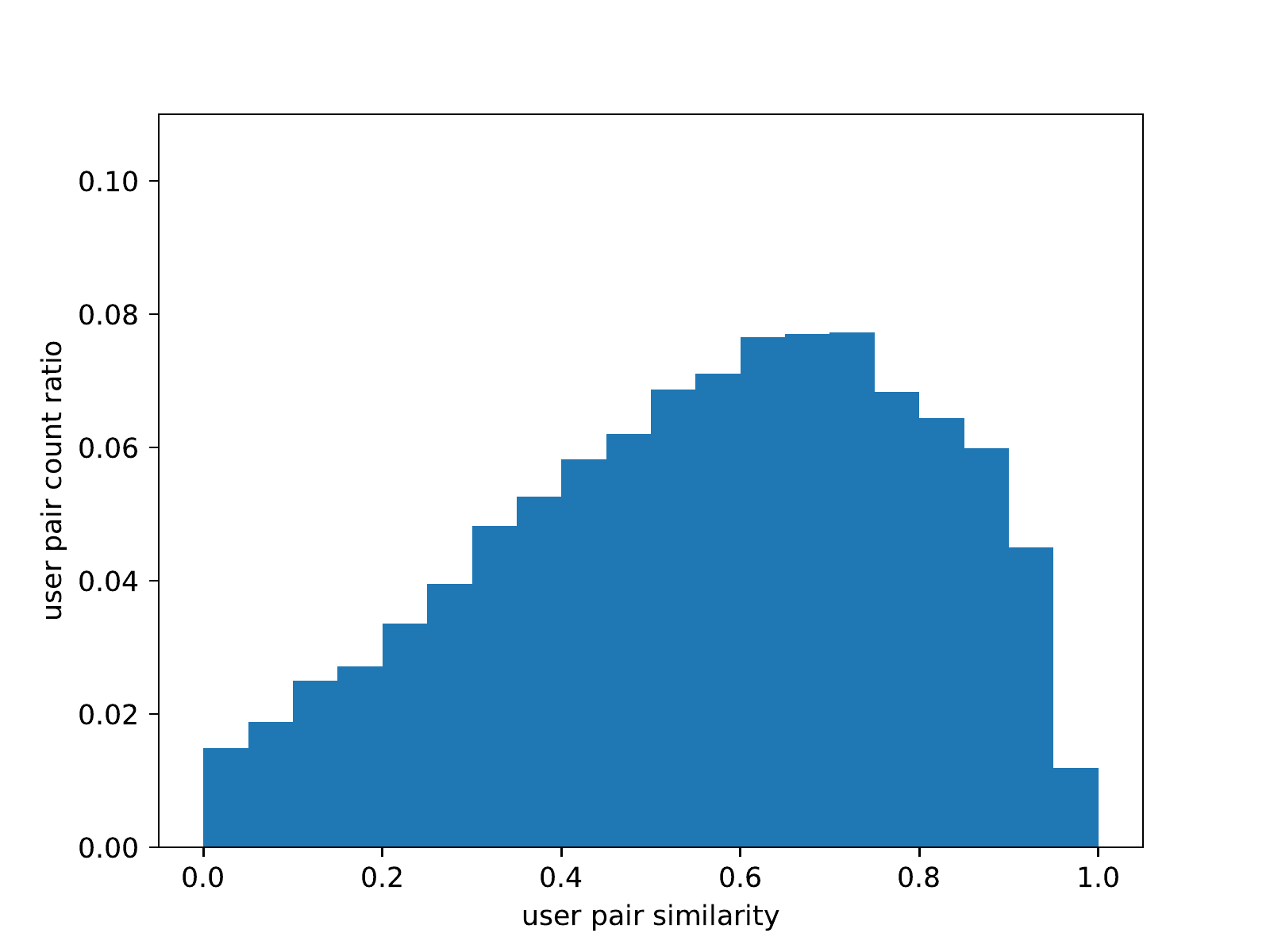}
    }
    \subfloat[\centering Friend (B) to Friend Hub (C)\label{fig:hist_B2C}]{
        \includegraphics[width=0.24\linewidth]{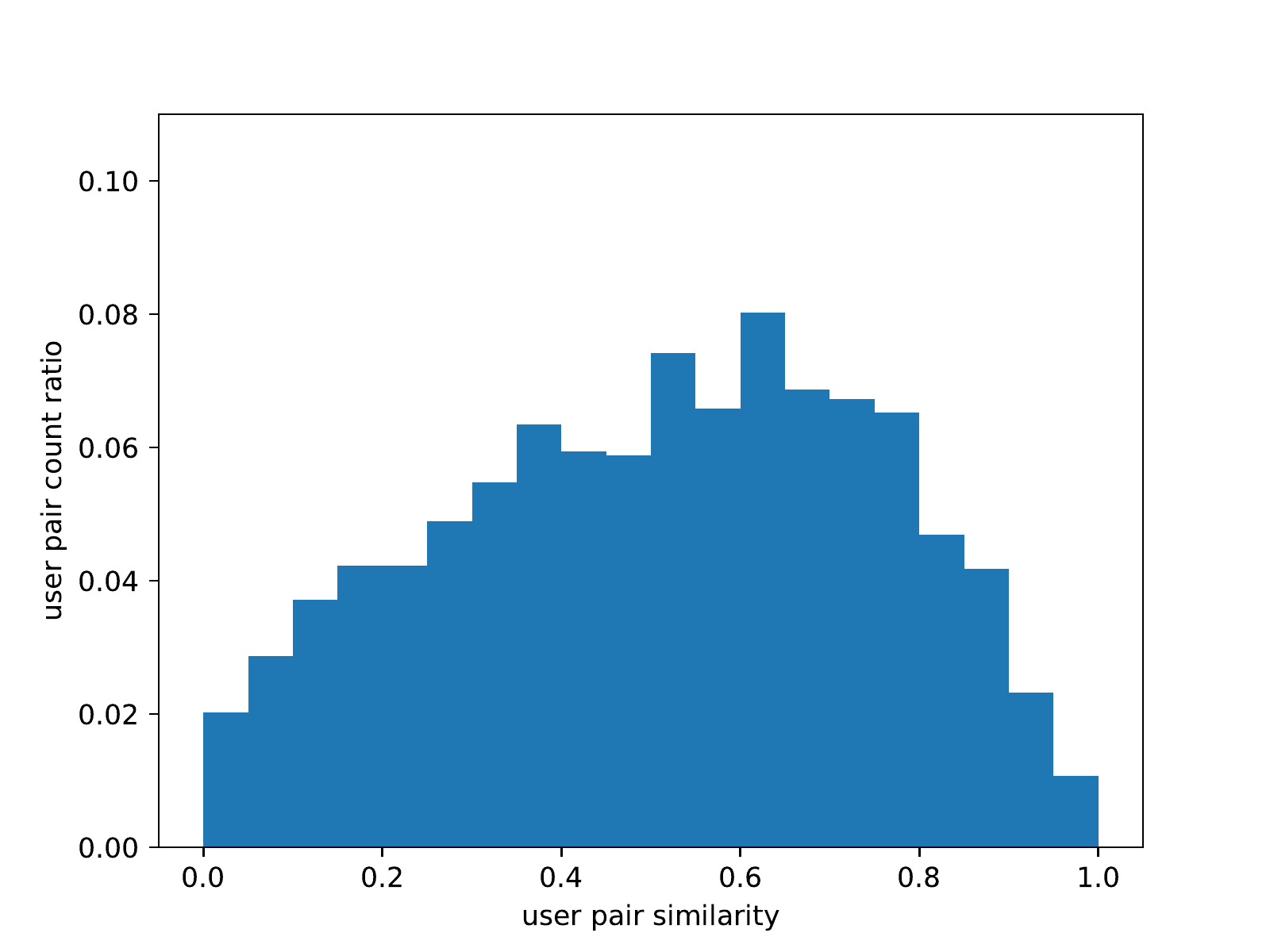}
    }
    \subfloat[\centering Friend (B) to\newline Information Source (D)\label{fig:hist_B2D}]{
        \includegraphics[width=0.24\linewidth]{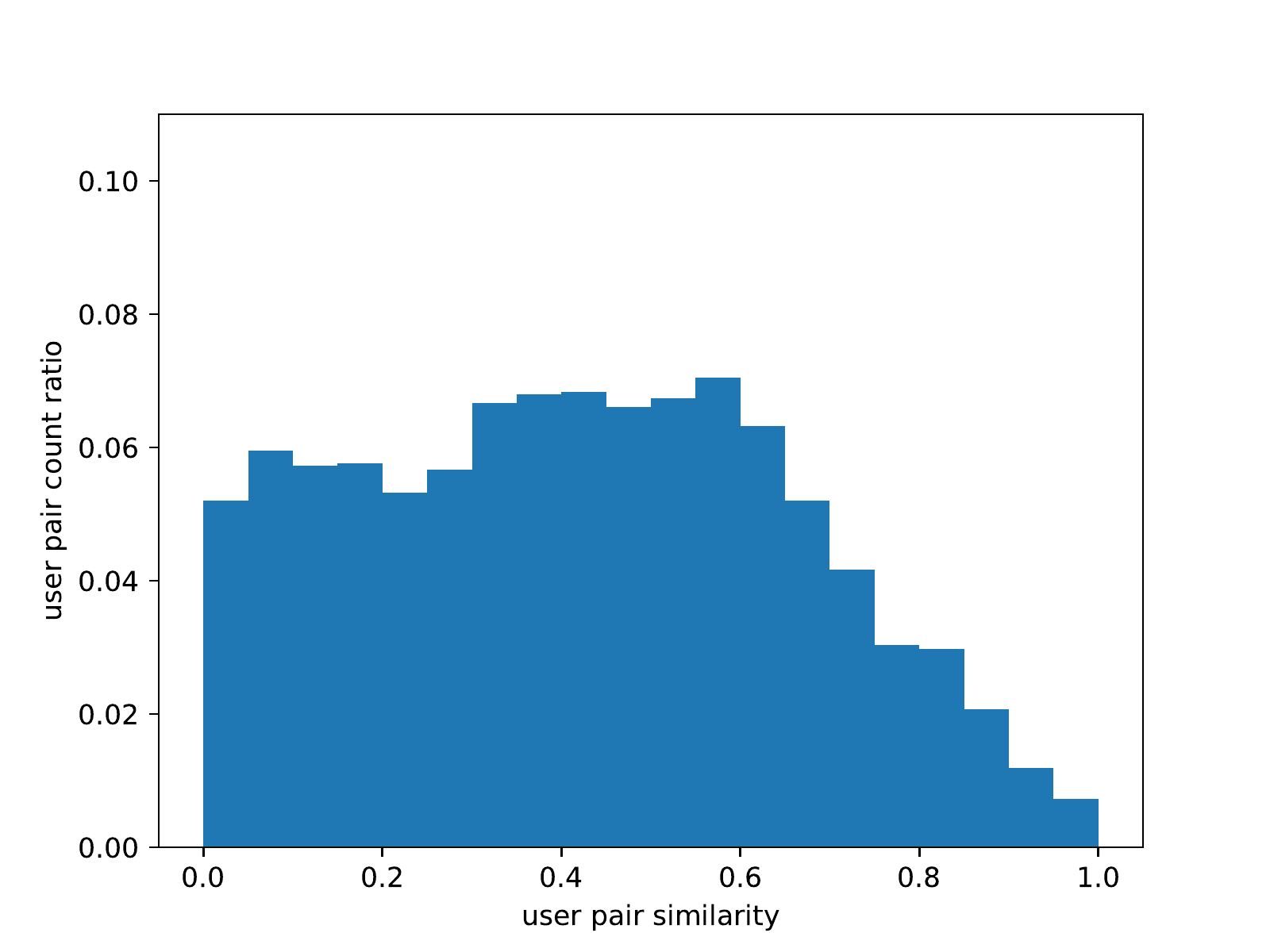}
    }\\
    \subfloat[\centering Friend Hub (C) to\newline Information Seeker (A)\label{fig:hist_C2A}]{
        \includegraphics[width=0.24\linewidth]{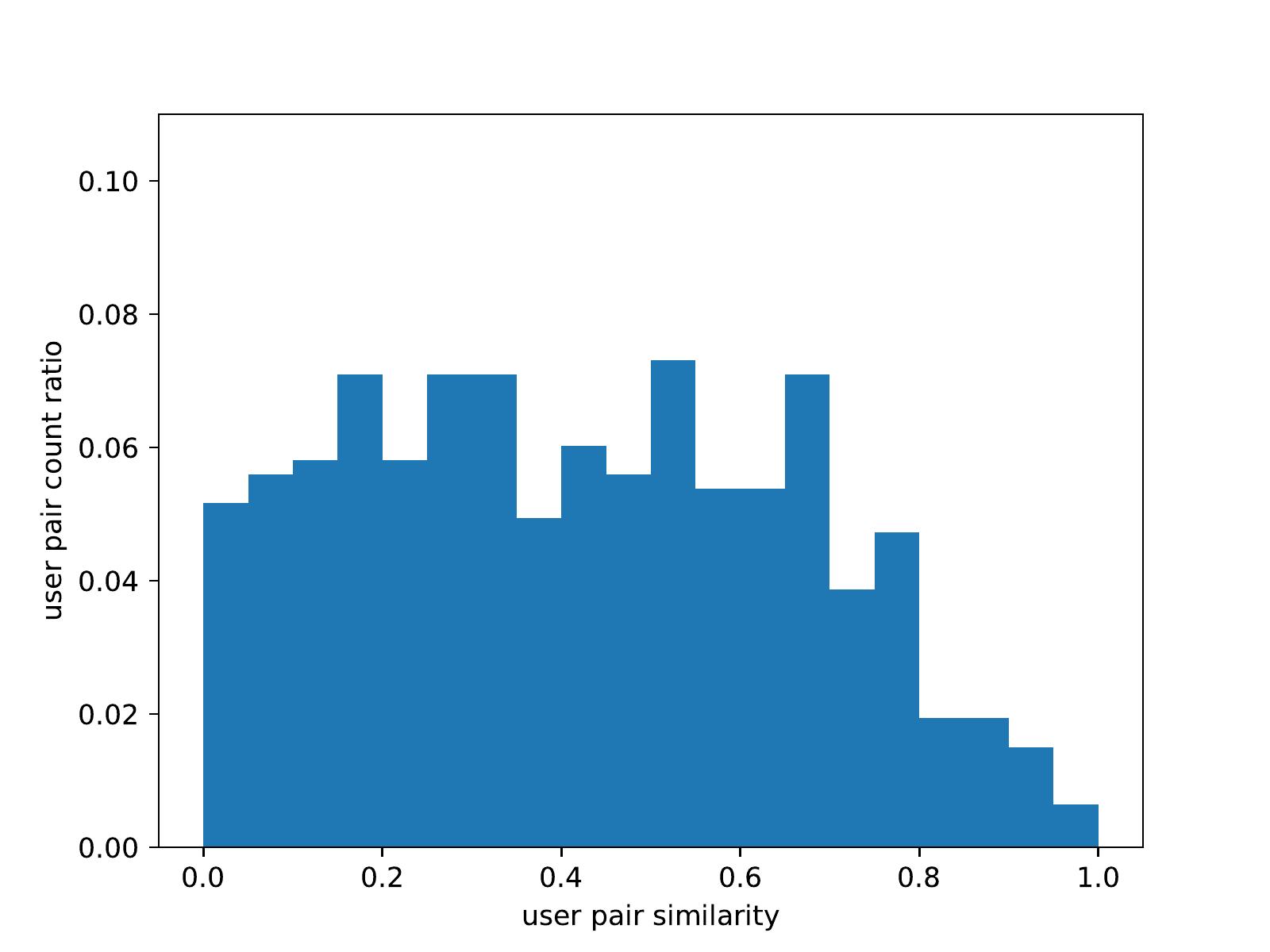}
    }
    \subfloat[\centering Friend Hub (C) to Friend (B)\label{fig:hist_C2B}]{
        \includegraphics[width=0.24\linewidth]{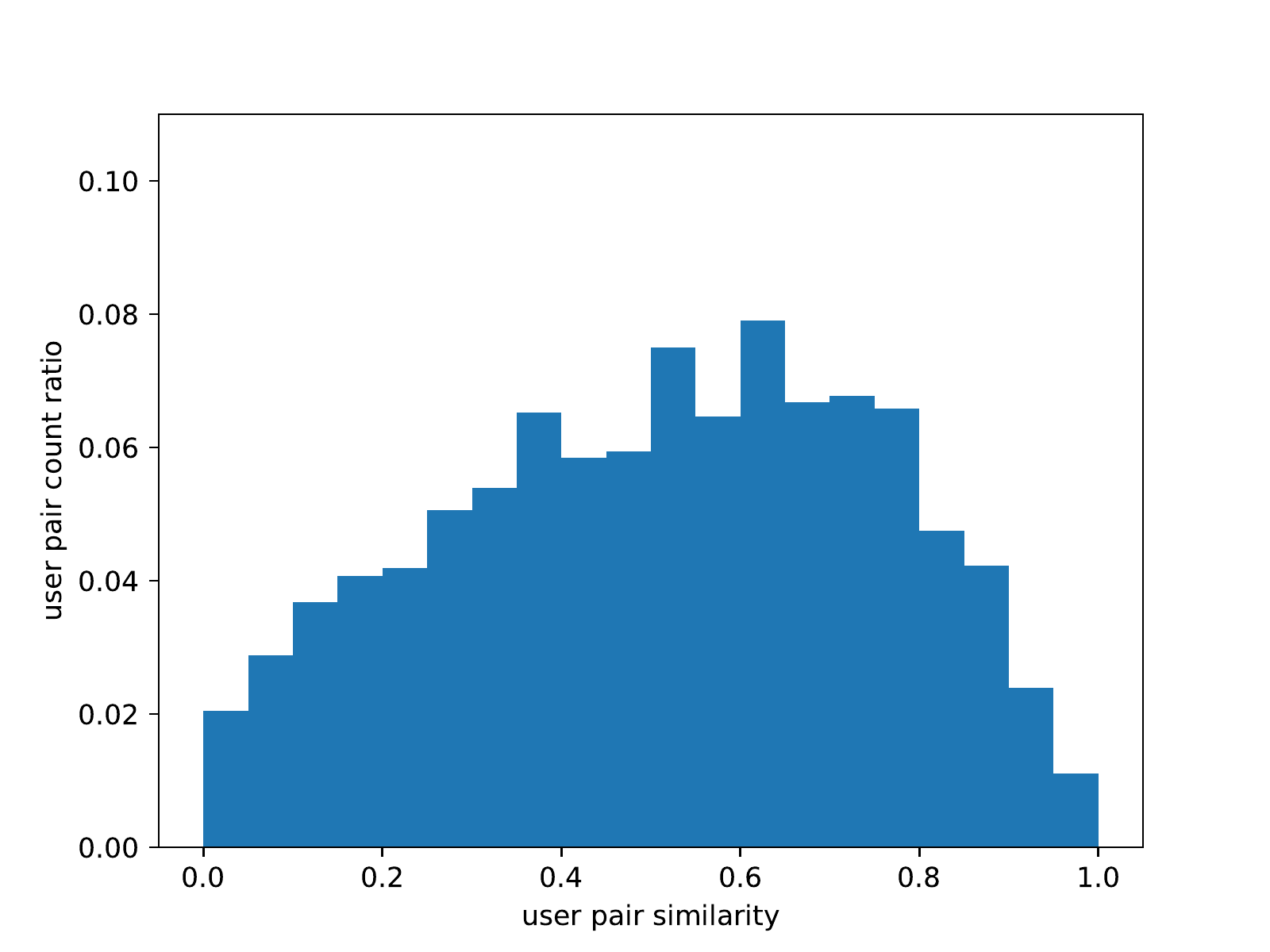}
    }
    \subfloat[\centering Friend Hub (C) to Friend Hub (C)\label{fig:hist_C2C}]{
        \includegraphics[width=0.24\linewidth]{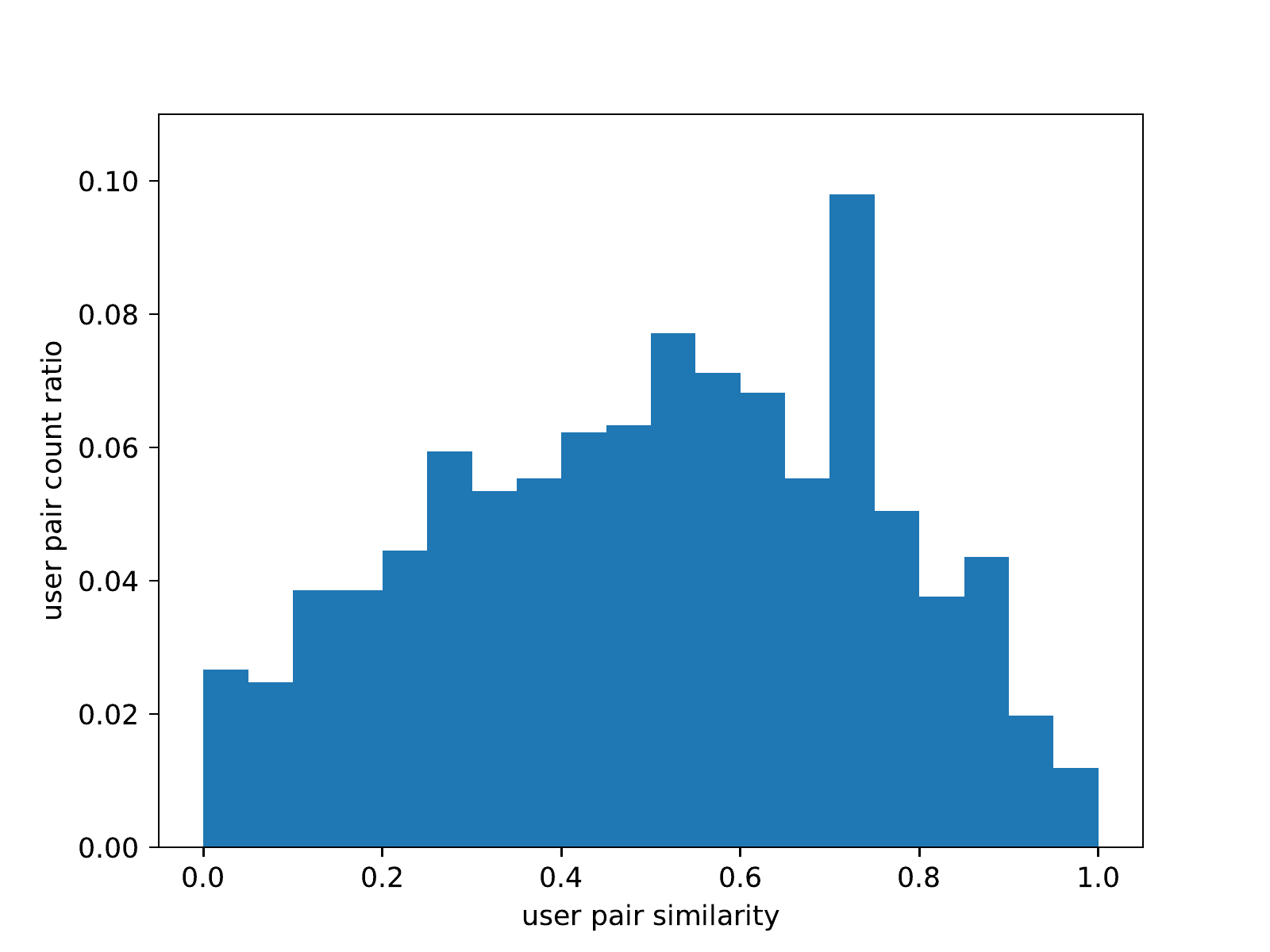}
    }
    \subfloat[\centering Friend Hub (C) to\newline Information Source (D)\label{fig:hist_C2D}]{
        \includegraphics[width=0.24\linewidth]{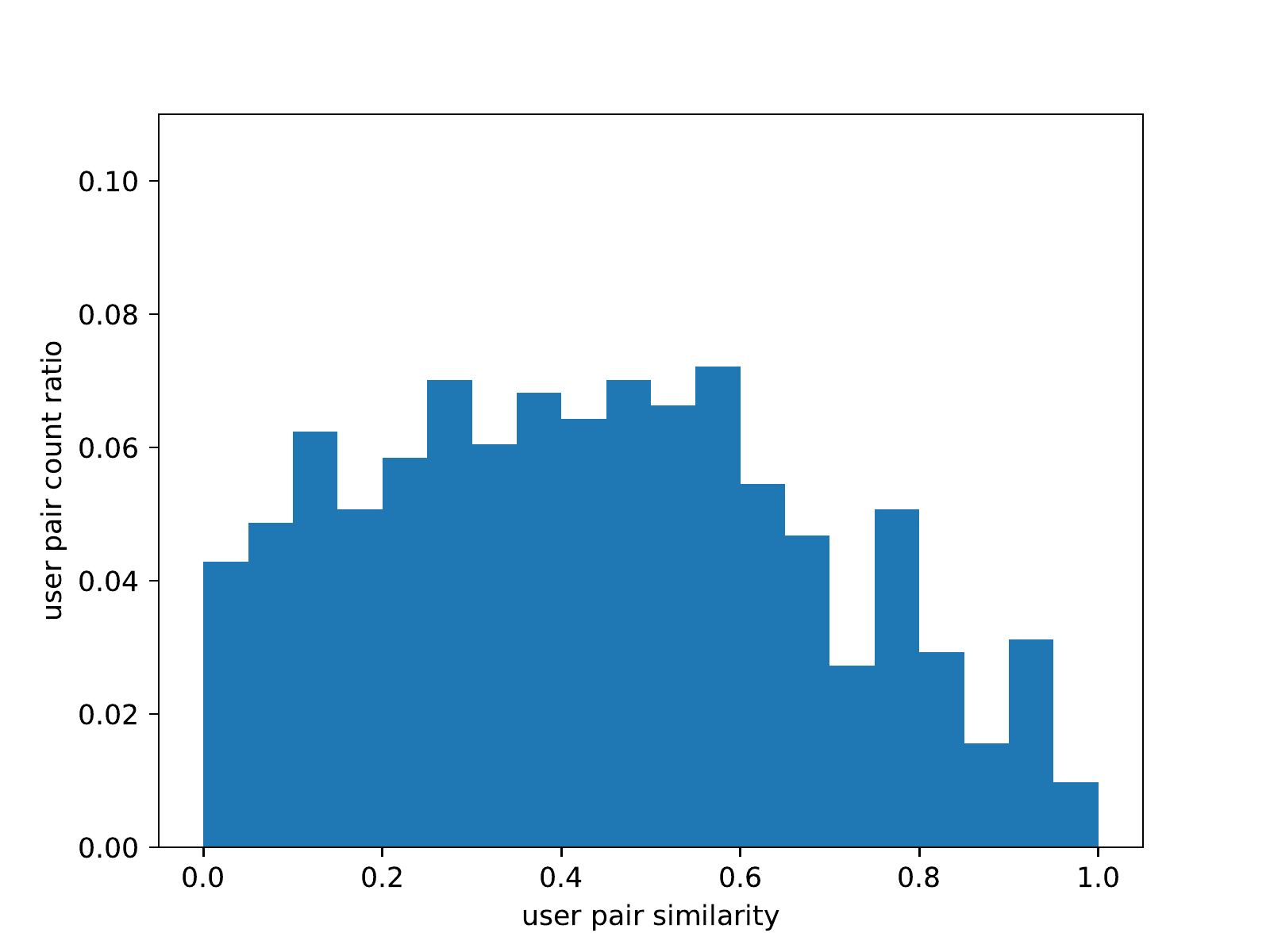}
    }\\
    \subfloat[\centering Information Source (D) to\newline Information Seeker (A)\label{fig:hist_D2A}]{
        \includegraphics[width=0.24\linewidth]{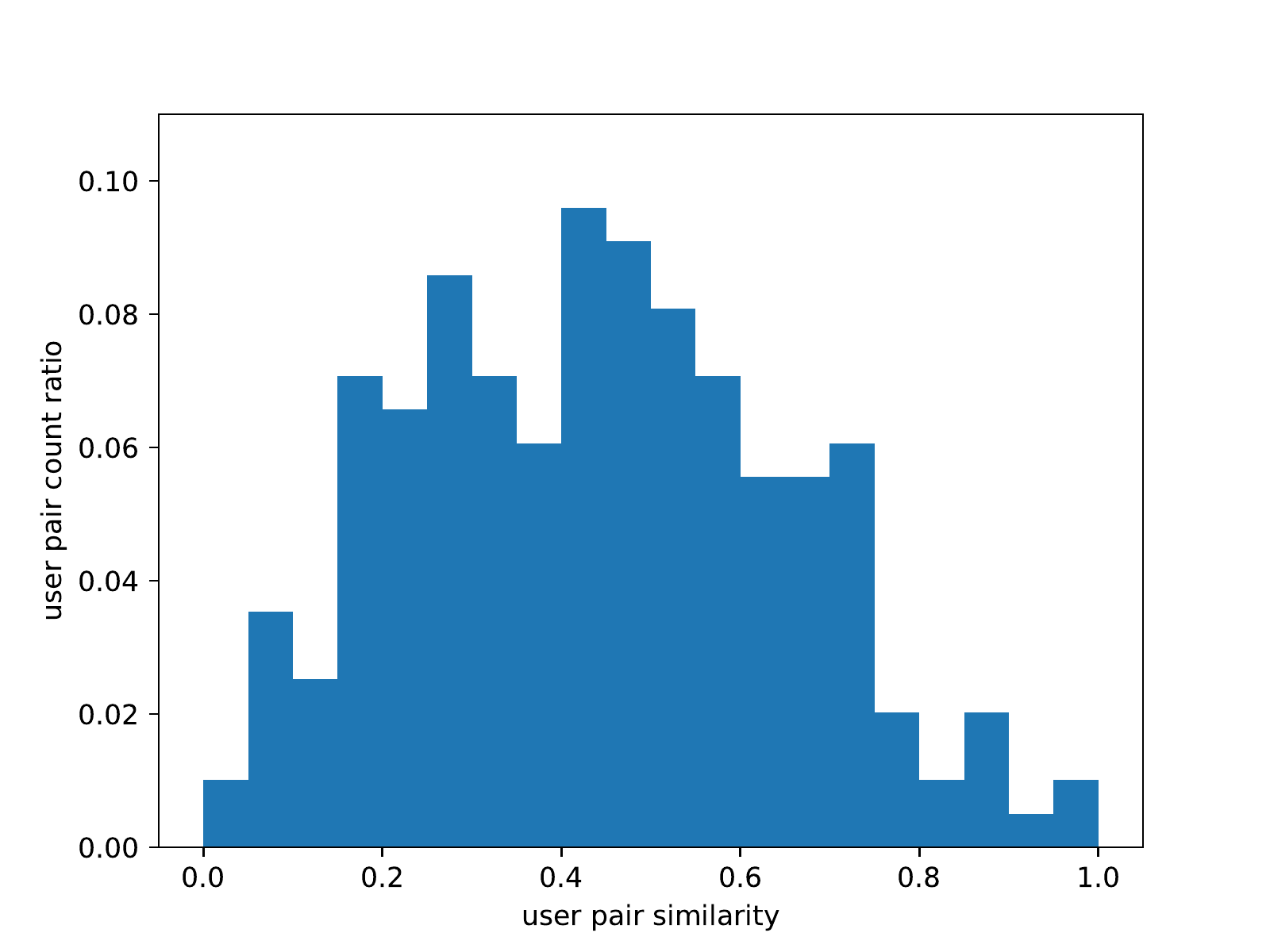}
    }
    \subfloat[\centering Information Source (D) to\newline Friend (B)\label{fig:hist_D2B}]{
        \includegraphics[width=0.24\linewidth]{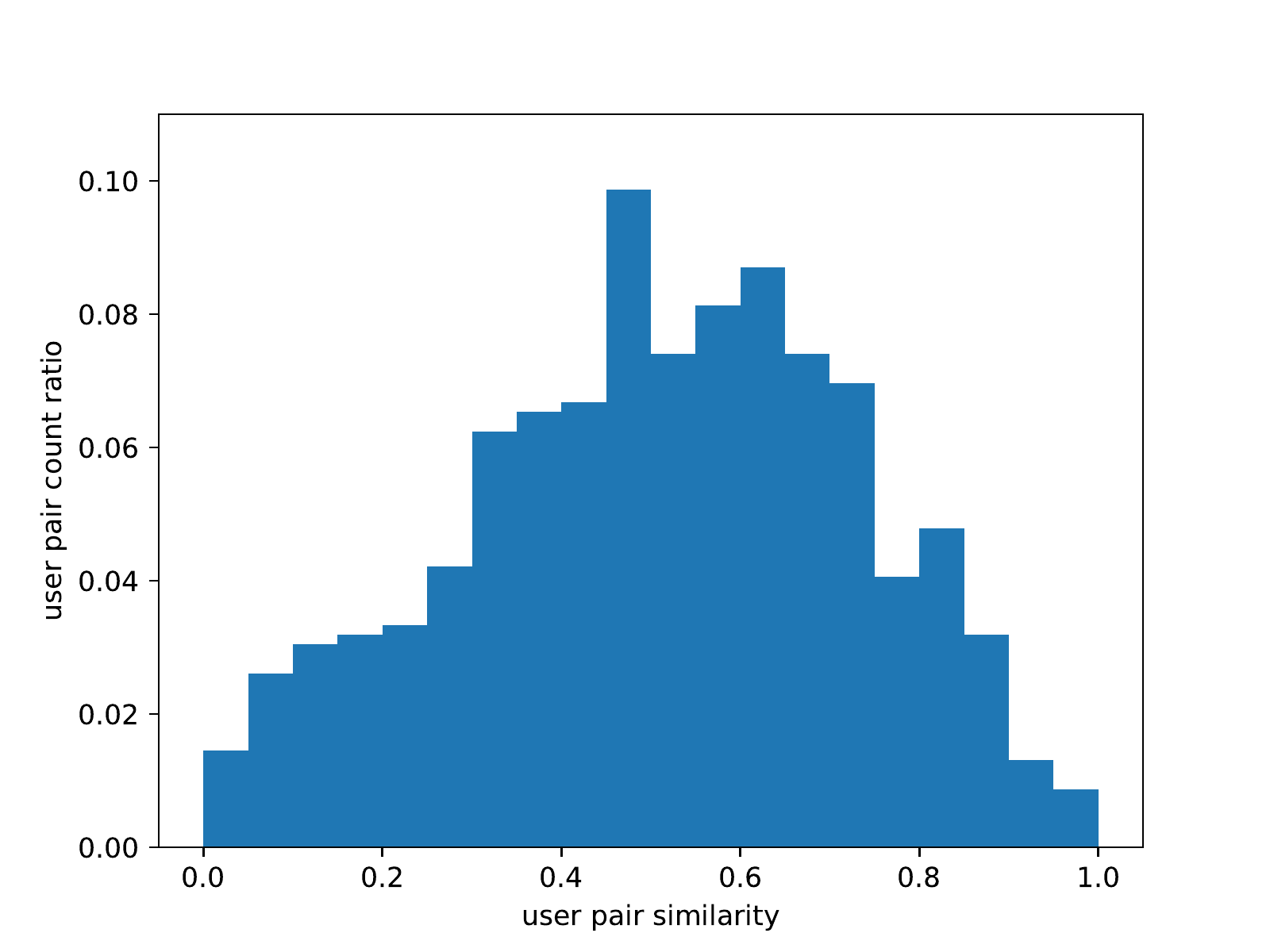}
    }
    \subfloat[\centering Information Source (D) to\newline Friend Hub (C)\label{fig:hist_D2C}]{
        \includegraphics[width=0.24\linewidth]{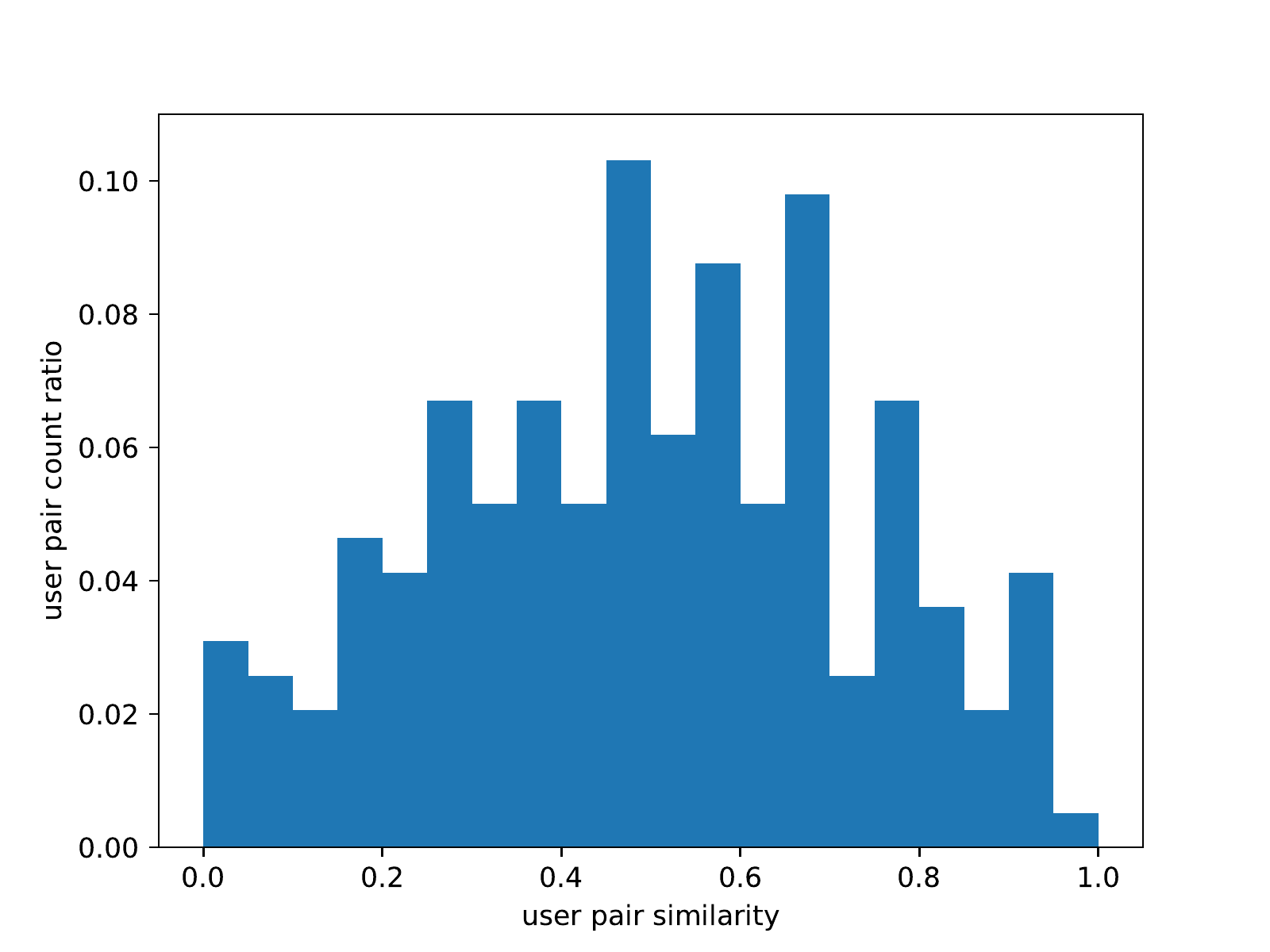}
    }
    \subfloat[\centering Information Source (D) to\newline Information Source (D)\label{fig:hist_D2D}]{
        \includegraphics[width=0.24\linewidth]{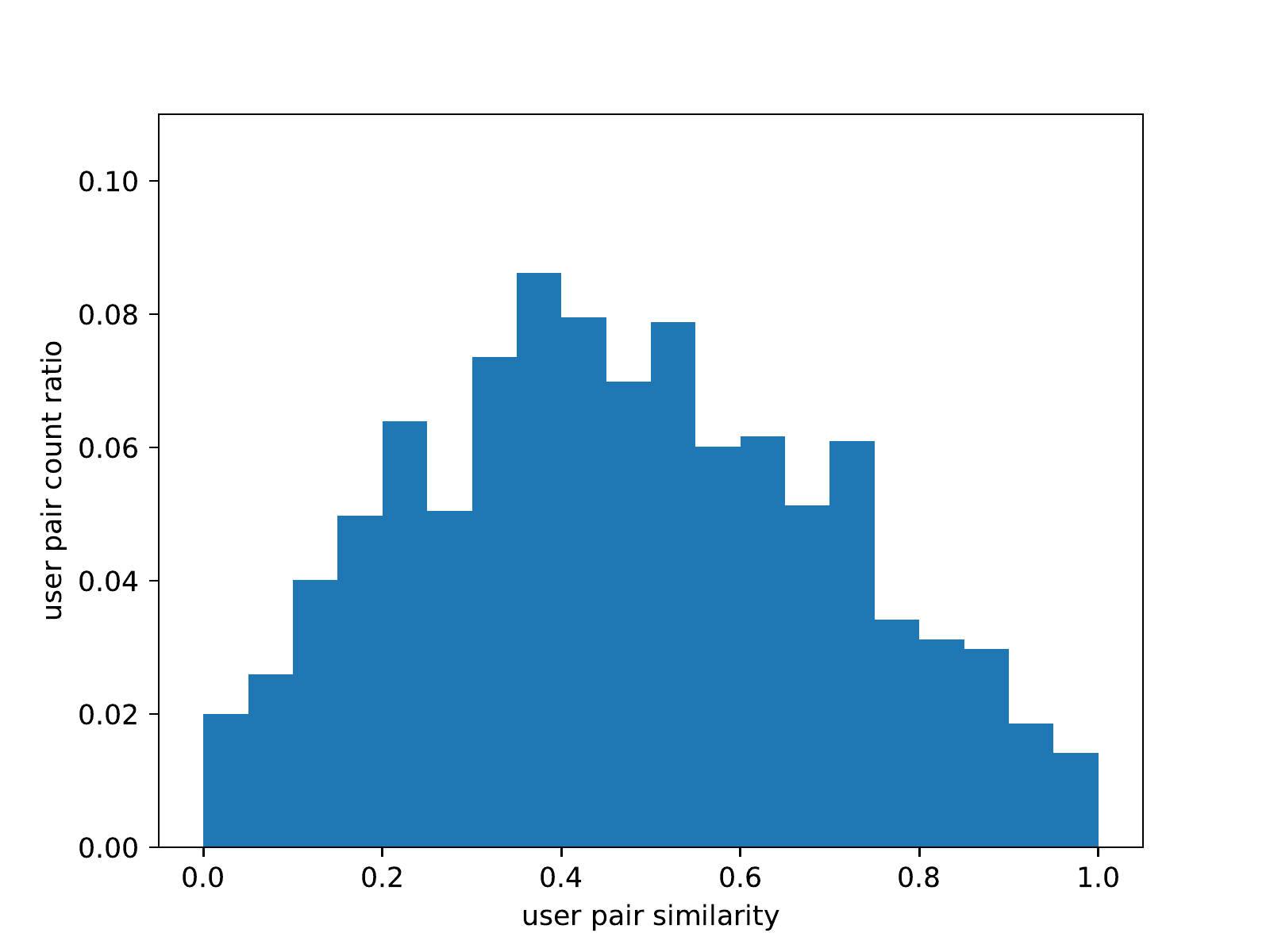}
    }
    \caption{Normalized histograms of similarity across user categories.}
    \label{fig:hist}
\end{figure*}

To compare the following preferences in topical homophily, we extracted the following relationships across categories and calculated their average similarity.
The average similarity between categories is shown in Table~\ref{tab:pair_sim_mean}.
The overall average similarity was 0.4412.
To examine this in more detail, the normalized histograms of similarity are shown in Figure~\ref{fig:hist}.
The number of histogram bins was set to 20.
Because only a few pairs had negative values, the minimum value on the x-axis was set to 0.
Based on the overall average similarity and histogram, we interpreted that a histogram with an upward trend indicated high topical similarity.

\begin{table}[t]
    \centering
    \caption{Average similarity across categories}
    \label{tab:pair_sim_mean}
    \begin{tabular}{|c|c|c|c|c|}\hline
        \multirow{2}{*}{row follows column} & Information & \multirow{2}{*}{Friend} & Friend & Information\\
        & Seeker && Hub & Source \\\hline
        Information& \multicolumn{1}{|r|}{\multirow{2}{*}{0.4643}} & \multicolumn{1}{|r|}{\multirow{2}{*}{0.4419}} & \multicolumn{1}{|r|}{\multirow{2}{*}{0.4167}} & \multicolumn{1}{|r|}{\multirow{2}{*}{0.3535}} \\
        Seeker&&&&\\\hline
        \multirow{2}{*}{Friend} & \multicolumn{1}{|r|}{\multirow{2}{*}{0.4462}} & \multicolumn{1}{|r|}{\multirow{2}{*}{0.5584}} & \multicolumn{1}{|r|}{\multirow{2}{*}{0.5061}} & \multicolumn{1}{|r|}{\multirow{2}{*}{0.4118}} \\
        &&&&\\\hline
        \multirow{2}{*}{Friend Hub} & \multicolumn{1}{|r|}{\multirow{2}{*}{0.4127}} & \multicolumn{1}{|r|}{\multirow{2}{*}{0.5070}} & \multicolumn{1}{|r|}{\multirow{2}{*}{0.5000}} & \multicolumn{1}{|r|}{\multirow{2}{*}{0.4255}} \\\
        &&&&\\\hline
        Information& \multicolumn{1}{|r|}{\multirow{2}{*}{0.4375}} & \multicolumn{1}{|r|}{\multirow{2}{*}{0.5085}} & \multicolumn{1}{|r|}{\multirow{2}{*}{0.4916}} & \multicolumn{1}{|r|}{\multirow{2}{*}{0.4647}} \\
        Source&&&&\\\hline
    \end{tabular}
\end{table}

First, we compared users belonging to Category B (Friend) and C (Friend Hub).
As shown in Figure~\ref{fig:hist}, the histograms between Categories B and C exhabited an upward trend.
Thus, the users in Categories B and C had high topical similarity.

Next, we focused on the similarities between users in Category D (Information Source) and their followers.
As shown in the fourth column of Figure~\ref{fig:hist}, these histograms exhabited a downward trend.
In particular, the histograms of the relationships from Categories A (Information Seeker) to D (Figure~\ref{fig:hist_A2D}) exhabited a strong downward trend.
Accordingly, users in Category D tended to be less similar to their followers.

\section{Discussion}
We discuss the characteristics of the users in Category A (Information Seeker).
We found that the users in Category A tended to follow the users in Category D (Information Source) (Table~\ref{tab:user_pairs}).
However, the similarity of the relationship across Categories A to D was low.
Therefore, the followings from Categories A to D were considered to be for information-gathering purposes, as the topics of their followees were less related to the other's topics and their own topics.

We assumed that the users in Categories B (Friend) and C (Friend Hub) had different characteristics.
The former had a follower--followee ratio near and above 1.0, and the latter had a follower--followee ratio near and below 1.0.
Because the follower--followee ratio was approximately 1, users in both categories were expected to many friendships and high topical similarity.
Calculating the percentage of followees' categories for each category from Table~\ref{tab:user_pairs}, we observed that the users in Categories B and C had a similar pattern, but Category B had more edges to Category D.
Thus, users in Categories B and C were similar but there were differences.
Compared the second and third rows of Figure~\ref{fig:hist}, the histograms had a similar shape.
This indicated that the users in Categories B and C had similar following preferences in terms of topical homophily.
Conversely, based on the similarity with followers (the second and third columns of Figure~\ref{fig:hist}), the histograms of Categories B and C were similar.
These results indicated that the tweet contents are similar between the users in Categories B and C, but the purpose of Twitter use may be different, such that Category B has more users with information-gathering purposes.

We discuss the relationship directed to the users in Category D (Information Source).
The users in Category D were maily followed by users in Category A (Information Seeker).
The users in Category B (Friend) also tended to follow the users in Category D.
However, because the users in Category D had low similarity with their followers, the users in Category D were followed for their information, not because they were friends.

\section{Conclusion}
In this study, we analyzed the following preferences of users based on user categories and topical similarity.
We generated user vectors using their tweets and confirmed the similarity between users with following relationship were higher than one between users without.
We focused on the follower--followee ratio, which represented the features of the network structure around a user, to classify users.
Using the ratio, we classified users into four categories.
Then, we counted the number of user pairs with following relationships for each category and calculated the average similarity between users with following relationships for each category.
As a result, we found that users in Category A (Information Seeker) preferred to follow users in Category D (Information Source).
The tweet contents of users in Categories B (Friend) and C (Friend Hub) were similar; however, the preference of following was notably different.
Users in Category D (Information Source) were followed by a wide range of users for their information.

The data that we used in this analysis can be collected automatically.
Futhermore the method to generate user vectors is scalable.
Therefore, our analysis method can be used for a large amount of data.
Because building a relationship with other users and posting text is a basic feature on social media platforms, our method can potentially be adapted to other social media platforms.

\bibliographystyle{IEEEtran}
\bibliography{references}
\end{document}